\documentclass[12pt]{article}

\usepackage[margin=1in]{geometry}
\usepackage{graphicx}
\usepackage{algorithm}
\usepackage[noend]{algpseudocode}
\usepackage{amsmath}
\usepackage{amssymb}
\usepackage{amsfonts}
\usepackage{latexsym}
\usepackage{pifont}
\usepackage{bbm}
\usepackage{hyperref}
\usepackage{mathrsfs}
\usepackage[nodisplayskipstretch]{setspace}
\usepackage{stackrel}
\usepackage{natbib}
\usepackage{xcolor}
\usepackage{appendix}
\usepackage{multirow}

\newtheorem{theorem}{Theorem}[section]

\newcommand{\qed}{\nobreak \ifvmode \relax \else 
      \ifdim\lastskip<1.5em \hskip-\lastskip
      \hskip1.5em plus0em minus0.5em \fi \nobreak
      \vrule height0.75em width0.5em depth0.25em\fi}

\def\bs{\boldsymbol}

\newcommand\blfootnote[1]{%
  \begingroup
  \renewcommand\thefootnote{}\footnote{#1}%
  \addtocounter{footnote}{-1}%
  \endgroup
}

\begin{document}

\title{Improved Pathwise Coordinate Descent for Power Penalties}
\author{Maryclare Griffin$^{1}$}
\date{August 14, 2023}

\maketitle

\blfootnote{\raggedright \hspace{-2.15em}
$^1$Department of Mathematics and Statistics, University of Massachusetts Amherst, Amherst, MA 01003 (\href{mailto:maryclaregri@umass.edu}{\texttt{maryclaregri@umass.edu}}). \\
This research was supported by NSF grant DMS-2113079. Replication code is available at \href{https://github.com/maryclare/powreg}{\texttt{https://github.com/maryclare/powreg}}.}

\begin{abstract} 
Pathwise coordinate descent algorithms have been used to compute entire solution paths for lasso and other penalized regression problems quickly with great success. They improve upon cold start algorithms by solving the problems that make up the solution path sequentially for an ordered set of tuning parameter values, instead of solving each problem separastely.
However, extending pathwise coordinate descent algorithms to more the general bridge or power family of $\ell_q$ penalties is challenging. Faster algorithms for computing solution paths for these penalties are needed because $\ell_q$ penalized regression problems can be nonconvex and especially burdensome to solve.
In this paper, we show that a reparameterization of $\ell_q$ penalized regression problems is more amenable to pathwise coordinate descent algorithms. This allows us to improve computation of the mode-thresholding function for $\ell_q$ penalized regression problems in practice and introduce two separate pathwise algorithms. We show that either pathwise algorithm is faster than the corresponding cold start alternative, and demonstrate that different pathwise algorithms may be more likely to reach better solutions.

\smallskip

\end{abstract}

\doublespacing
\newpage

\section{Introduction}

Consider the problem of computing regression coefficients subject to an $\ell_q$ penalty, sometimes called a bridge or power penalty, \color{black} which minimizes \color{black}
\begin{align}\label{eq:alg}
\frac{1}{2}\left|\left|\bs y - \bs X \bs \beta\right|\right|^2_2 + \lambda \left|\left|\bs \beta\right|\right|^q_q
\end{align}
\color{black}with respect to $\bs \beta$, \color{black} where $\bs y$ is an $n\times 1$ response vector, $\bs X$ is an $n\times p$ matrix of covariates, $\bs \beta$ is a $p\times 1$ vector of regression coefficients, and $\lambda$ is a tuning parameter \citep{Frank1993}.
The $\ell_q$ penalty includes the $\ell_2$ ridge, $\ell_0$ best subset, and $\ell_1$ lasso penalties as special cases. \color{black} 
When $q < 1$ the $\ell_q$ penalty is nonconvex and multiple minimizers of \eqref{eq:alg} may exist, however $\ell_q$ penalties are nonetheless valued because they can yield \color{black} sparser solutions with less bias \citep{Huang2008}. \color{black}
\color{black}
It can be difficult to compute a value of $\bs \beta$ that \color{black}minimizes \color{black} \eqref{eq:alg}, especially when $q < 1$ \citep{Mazumder2011}. 
\color{black}
This is often magnified by the need to \color{black}find minimizers of \color{black} \eqref{eq:alg} for a collection of values of $\lambda$ and $q$, because a single optimal choice of $\lambda$ and $q$ is rarely known and often data dependent \citep{Griffin2017c}. 

Pathwise coordinate descent algorithms have been one popular approach to overcoming the computational challenges encountered when solving penalized regression problems. 
\color{black}
Coordinate descent algorithms provide a method for solving \eqref{eq:alg} for fixed $\lambda$ and $q$ that compute a solution to \eqref{eq:alg} by iteratively minimizing with respect to one coordinate of $\bs \beta$ at a time, holding the rest fixed. This corresponds to iteratively computing the mode-thresholding function, $g\left(\lambda, q; b\right)$, which minimizes
\begin{align}\label{eq:mold}
\frac{1}{2}\left(b - \beta\right)^2 + \lambda \left|\beta\right|^q
\end{align} 
with respect to $\beta$, where $\beta$ corresponds to the coordinate $\beta_j$ that \eqref{eq:alg} is being maximized with respect to and the values of $b$ and $\lambda$ are determined by the data and the remaining coordinates of $\bs \beta$.  Pathwise algorithms build on methods that solve \eqref{eq:alg} for fixed $\lambda$ and $q$ to solve \eqref{eq:alg} for a collection of values of $\lambda$ and $q$. Although not inextricably linked, pathwise algorithms are often built on a foundation of coordinate descent algorithms.
\color{black}

\color{black}
Coordinate descent algorithms which solve \eqref{eq:alg}  by iteratively solving \eqref{eq:mold} are provided in in \cite{Fu1998} for $q \geq 1$ and \cite{Marjanovic2014} for $q < 1$. \cite{Marjanovic2014} also provide conditions on $\bs X$ that guarantee convergence to a local optimum and verifiable conditions for local optimality of a coordinate descent solution. \cite{Fu1998} provides verifiable optimality conditions for $q \geq 1$. In the case of lasso penalized regression, which corresponds to the special case of \eqref{eq:alg} when $q = 1$, pathwise coordinate descent methods have been developed \citep{Friedman2007}.
\color{black}
These pathwise coordinate descent algorithms solve \eqref{eq:alg}  for a specific value of the tuning parameter $\lambda^*$ by finding a value $\lambda_0$ that ensures that the minimizing $\bs \beta$ is exactly equal to zero, and then solving \color{black}\eqref{eq:alg} along a path of tuning parameter values $\lambda_0, \dots, \lambda^*$ using coordinate descent, using the minimizing $\bs \beta$ for one problem as a starting value for the next. This works well when $q = 1$ because $\lambda_0$ is easy to determine from the data, and because mode-thresholding function, $f\left(\lambda; b \right) = \text{argmin}_{\beta} \left(b - \beta\right)^2 + \lambda \left|\beta\right|$,
is nested in $\lambda$, i.e. if $f\left(\lambda'; b \right) = 0$ and  $\epsilon > 0$, then  $f\left(\lambda' + \epsilon; b \right) = 0$.  \color{black}

\begin{figure}
	\centering
	\includegraphics{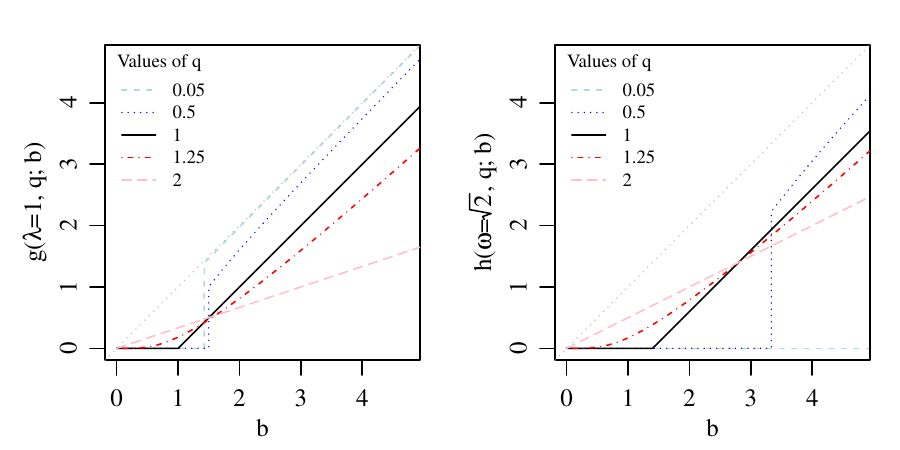}
	\caption{The left panel shows the thresholding function for the parametrization considered in \cite{Mazumder2011}. The right panel shows the thresholding function for the parametrization given in Equation~\eqref{eq:mtfre}.}
	\label{fig:thresh}
\end{figure}

Unfortunately, pathwise coordinate descent algorithms have been more challenging to develop for $\ell_q$ penalties \citep{Mazumder2011}.
There are \color{black}three \color{black} main challenges that arise in the development of pathwise coordinate descent algorithms for the bridge/power family of penalties, \color{black} all \color{black} of which are related to properties of the mode-thresholding function, \color{black} $g\left(\lambda, q; b\right)$. \color{black}
First, the mode-thresholding function is only available in closed form in special cases. When $q = 1$, $h\left(\lambda, q = 1; b\right) = \text{sign}\left(b\right)\left(\left|b\right| - \lambda \right)_+$, where $\left(x\right)_+ = \text{max}\left\{x, 0\right\}$. When $q = 2$, $h\left(\lambda, q = 2; b\right) = \left( 1+ 2\lambda \right)^{-1}b$.
Otherwise, a closed-form solution is not available. This means that \color{black}initializing \color{black} a pathwise coordinate descent algorithm for fixed values of $q < 1$ \color{black}by finding \color{black} a value of the tuning parameter $\lambda_0$ that ensures that the solution to \eqref{eq:alg} is exactly equal to the zero vector \color{black} can require additional iterative computation, which is possible but can be inconvenient in practice. \color{black} 
\color{black} Second, the mode-thresholding function has two solutions at a single value of $b$ when $q < 1$. \color{black}
\color{black}Third, \color{black} the mode-thresholding function is not nested in $q$ for fixed $\lambda$, i.e. if $g\left(\lambda', q'; b\right) = 0$ and $0 \leq \epsilon < q'$, it may not be true that $g\left(\lambda', q' - \epsilon; b\right) = 0$. 
This can be observed in the left panel of Figure~\ref{fig:thresh}, in which the tuning parameter $\lambda$ is fixed at $1$. For a small range of values of $b \approx 1.5$, the mode-thresholding function returns zero when $q = 0.5$ but not when $q = 0.05$.
\color{black} This is counterintuitive. As $q$ decreases, the corresponding penalty is expected to encourage sparsity more aggressively.\color{black}

The first \color{black} two challenges have been addressed by \cite{Marjanovic2014}, who introduced a coordinate descent algorithm for solving \eqref{eq:alg} for $q < 1$ which includes a method for  computing the mode-thesholding function $g\left(\lambda, q, b\right)$, \color{black} instructions for choosing a single solution to the mode-thresholding function when multiple exist in the context of coordinate descent, and \color{black} conditions for when the mode-thresholding function $g\left(\lambda, q, b\right) = 0$. \color{black} In this technical note, we show that a simple reparametrization of \eqref{eq:alg} eliminates the second challenge for $0 < q < 2$,
\begin{align}\label{eq:algre}
\frac{1}{2}\left|\left|\bs y - \bs X \bs \beta\right|\right|^2_2 + \left(\frac{\omega^{2-q}}{q}\right) \left|\left|\bs \beta\right|\right|^q_q,
\end{align}
in which the tuning parameter $\lambda$ is replaced by the tuning parameter $\omega$. 
\color{black} We emphasize that we use this reparameterization to provide new pathwise coordinate descent algorithms with desirable properties that can make use of existing coordinate descent algorithms for fixed $q$ and $\omega$, as opposed to new coordinate descent algorithms for solving \eqref{eq:alg} for fixed $q$ and $\omega$. \color{black}

The right panel of Figure~\ref{fig:thresh} shows the mode-thresholding function corresponding to \eqref{eq:algre}, \color{black} $h\left(\omega, q; b\right)$, which minimizes \color{black}
\begin{align}\label{eq:mtfre}
\frac{1}{2}\left(b - \beta\right)^2 +  \left(\frac{\omega^{2-q}}{q}\right) \left|\beta\right|^q.
\end{align}
In Figure~\ref{fig:thresh}, the mode-thresholding function appears to be nested in $q$ for fixed $\omega$, i.e. if $h\left(\omega', q'; b\right) = 0$ and $0 \leq \epsilon < q' \leq 1$,  then $h\left(\omega', q' - \epsilon; b\right) = 0$. Furthermore, for fixed $\omega$ there appears to be a unique value $b^*_{\omega}$ of $b$ for which all nonzero values of the mode-thresholding function are equal regardless of $q$ and for which nonzero values of the mode-thresholding function are increasing in $q$ for all $b < b^*_{\omega}$ and decreasing in $q$ for all $b > b^*_{\omega}$.

In what follows, we derive properties of the mode-thresholding function $h\left(\omega, q; b\right)$. Specifically, we derive a minimum value $\omega_q$ which satisfies $h\left(\omega, q; b\right) = 0$ for all $\omega > \omega_q$. We also prove that $h\left(\omega, q; b\right)$ is nested in $\omega$ for fixed $q$ and derive $b^*_\omega$, the value of $b$ for which all nonzero values of the mode-thresholding function are equal regardless of $q$. Last, we prove that sparsity of $h\left(\omega, q; b\right)$ is nested in $q$ for fixed $\omega$ that the nonzero values of $h\left(\omega, q; b\right)$ are increasing in $q$ for $b < b^*_{\omega}$ and decreasing in $q$ for $b > b^*_{\omega}$. All proofs are provided in an appendix. We then show how this knowledge can be used to improve computation of $h\left(\omega, q; b\right)$ in practice and introduce two different pathwise coordinate descent algorithms for solving \eqref{eq:algre}. One is similar to the pathwise coordinate descent algorithm for solving the lasso penalized regression problem, insofar as it computes a sequence of solutions for fixed $q$ starting from the a value of $\omega$ that yields an optimal value of $\bs \beta$ that is exactly equal to the zero vector for $q \leq 1$. The other computes a sequence of solutions for fixed $\omega$ starting from $q = 2$. Because \eqref{eq:algre} is nonconvex when $q < 1$, we compare not only timing of the two algorithms relative to their cold start alternatives and each other, but also the potential for each algorithm to reach a better mode.

\section{Properties and New Pathwise Algorithms}

Combining the results from \cite{Marjanovic2014} for $q < 1$ with what is known for $1 \leq q \leq 2$, the mode-thresholding function $h\left(\omega, q; b \right)$ defined in \eqref{eq:mtfre} satisfies, 
\begin{align}
h\left(\omega, q; b \right) = \left\{ \begin{array}{cc} 
0 & q \leq 1 \text{ and } \left|b\right| < \alpha\left(\omega, q\right) \\
\left\{0, \text{sign}\left(b\right)\gamma\left(\omega, q\right)\right\}  & q \leq 1 \text{ and } \left|b\right| = \alpha \left(\omega, q\right) \\
\text{sign}\left(b\right)\phi\left(\omega, q\right) & q > 1 \text{ or } q \leq 1 \text{ and } \left|b\right| > \alpha \left(\omega, q\right),
\end{array}\right. \label{eq:hre} 
\end{align}
where 
\begin{align}
\gamma\left(\omega, q\right) &= \left(\frac{1}{\omega}\right)\left(2 \left(\frac{1 - q}{q}\right) \right)^{\frac{1}{2-q}} \label{eq:gammare} \\
\alpha\left(\omega, q\right) &= \omega \left(2\left(1 - q\right)\right)^{\frac{q - 1}{2-q}}\left(2 - q\right) q^{\frac{1}{q-2}} \label{eq:alphare}
\end{align}
and $\phi\left(\omega, q\right) > 0$ is the larger of at most two values that satisfy
\begin{align}\label{eq:phire}
\phi\left(\omega, q\right) + \omega \left(\frac{\phi\left(\omega, q\right)}{\omega}\right)^{q-1} = \left|b\right|
\end{align}
when $ q > 1$ or $q \leq 1$ and $\left|b\right| > \alpha \left(\omega, q\right)$. These properties allow us obtain a condition for $\omega$ that ensures $h\left(\omega, q; b\right) = 0$ when $q\leq 1$, 
\begin{align}\label{eq:maxomega}
\omega  > \left(\frac{\left|b\right|}{2-q}\right) \left(2\left(1 - q\right)\right)^{\frac{1 - q}{2-q}}q^{\frac{1}{2-q}}.
\end{align} 

 \begin{figure}
	\centering
	\includegraphics{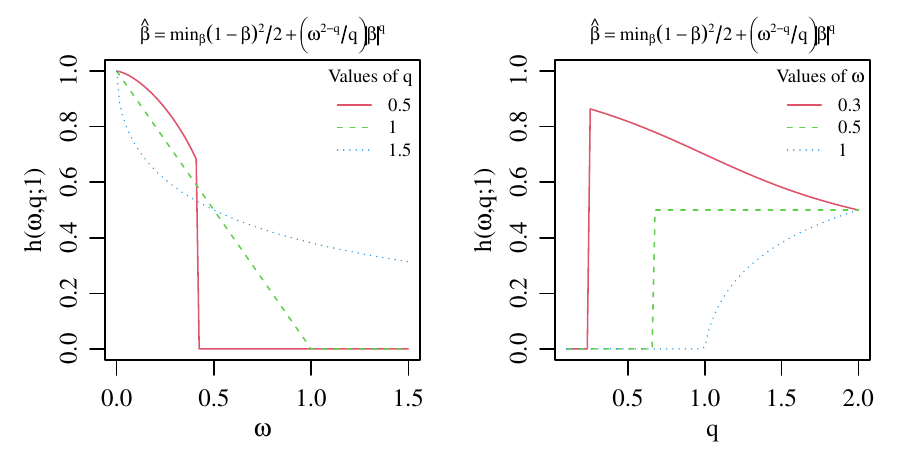}
	\caption{The left panel shows the thresholding function for the parametrization given in Equation~\eqref{eq:mtfre} as a function of $\omega$ for fixed $q$, the right panel as a function of $q$ for fixed $\omega$.}
	\label{fig:threshtwo}
\end{figure}

As suggested by Figure~\ref{fig:threshtwo}, the mode-thresholding function $h\left(\omega, q; b \right)$ is nested in $\omega$ and $q$. Proofs of the following specific claims are provided in an appendix.
Focusing on the left panel of Figure~\ref{fig:threshtwo}, we see evidence that sparsity of $h\left(\omega, q; b \right)$ is nested in $\omega$ for fixed $q \leq 1$.

\begin{theorem}
For fixed $q$ and $h\left(\omega, q; b\right)$ defined in Equation~\eqref{eq:hre}, if $h\left(\omega, q; b\right) = 0$ and $\omega' > \omega$ then $h\left(\omega', q; b\right) = 0$.
\end{theorem}

The left panel of Figure~\ref{fig:threshtwo} also suggests that all nonzero mode-thresholding function values for fixed $\omega$ are equal when $\omega = \left|b\right|/2$, with $\phi\left(\omega = \left|b\right|/2, q\right)  =  \left|b\right|/2$. 

\begin{theorem}\label{theorem:samephi}
For $h\left(\omega, q; b\right)$ defined in Equation~\eqref{eq:hre} and $q$ satisfying $q > 1$ or $q \leq 1$ and 
\begin{align}\label{eq:qcond}
 \left(2/\left(2-q\right)\right) \left(2\left(1 - q\right)\right)^{\frac{1 - q}{2-q}}q^{\frac{1}{2-q}} > 1,
\end{align}
 $h\left(\omega = \left|b\right|/2, q; b\right) = b/2$.
\end{theorem}

The left panel of Figure~\ref{fig:threshtwo} also suggests that nonzero values of $h\left(\omega, q; b \right)$ are nested in $\omega$ for fixed $q$.
\begin{theorem}\label{theorem:decomega}
For fixed $q$, if $\left|h\left(\omega, q; b\right)\right| > 0$ and $\omega' < \omega$ then  $\left|h\left(\omega', q; b\right)\right| \geq \left|h\left(\omega, q; b\right)\right| > 0$.
\end{theorem}

Turning to the right panel of Figure~\ref{fig:threshtwo}, we see evidence that sparsity of $h\left(\omega, q; b \right)$ is nested in $q$ for fixed $\omega$.

\begin{theorem}
For fixed $\omega$ and $h\left(\omega, q; b\right)$ defined in Equation~\eqref{eq:hre}, if $h\left(\omega, q; b\right) = 0$ and $q' < q \leq 1$ then $h\left(\omega, q'; b\right) = 0$.
\end{theorem}

The right panel of Figure~\ref{fig:threshtwo} also shows evidence that nonzero values of $h\left(\omega, q; b \right)$ are nested in $q$ for fixed $\omega$.

 \begin{theorem} 
For fixed $\omega > 0$, define 
 \begin{align*}
 \tilde{q}_{\omega/\left|b\right|} = \text{inf} \left\{ q: q = 1 \text{ or } q < 1 \text{ and } \left(\frac{1}{2-q}\right) \left(2\left(1 - q\right)\right)^{\frac{1 - q}{2-q}}q^{\frac{1}{2-q}} > \omega/\left|b\right| \right\}.
\end{align*}
If $\omega < \left|b\right|/2$ and $q' > q >  \tilde{q}_{\omega/\left|b\right|}$, then $\left|h\left(\omega, q; b\right)\right| \geq \left|h\left(\omega, q'; b\right)\right| > 0$. If $\omega > \left|b\right|/2$ and $q' > q > \tilde{q}_{\omega/\left|b\right|}$, then $\left|h\left(\omega, q'; b\right)\right| \geq \left|h\left(\omega, q; b\right)\right| > 0$.
\end{theorem}

These properties motivate two pathwise coordinate descent algorithms. The first computes a sequence of solutions for fixed $q$ and varying values of $\omega$. 

\begin{algorithm}[H]
\begin{enumerate}
	\item Based on \eqref{eq:maxomega}, define 
	\begin{align*}
	\omega^{\left(min\right)}_q = \left\{\begin{array}{cc}
	\text{max}_j\left(\boldsymbol x_j'\boldsymbol x_j\right)^{\frac{q-1}{2-q}}\left(\frac{\left|\boldsymbol x_j'\boldsymbol y\right|}{2-q}\right) \left(2\left(1 - q\right)\right)^{\frac{1 - q}{2-q}}q^{\frac{1}{2-q}}
	& \text{ if } q\leq 1 \\
	\text{max}_j \left|\boldsymbol x_j'\boldsymbol y\right| & \text{otherwise.}
	\end{array}\right. 
	\end{align*} 
	For $q \leq 1$, this is the smallest value of $\omega$ that yields an exactly zero solution for $\boldsymbol \beta$. 
	\item Fix a sequence of $k_{\omega}$ strictly decreasing $\omega$ values, $\boldsymbol \omega = \left\{\omega_1,\dots, \omega_{k_{\omega}}\right\}$ where $\omega_1 = \omega^{\left(min\right)}_q$. Let $\bs B$ refer to the $p\times k_{\omega}$ array of solutions.
	\item Solve for $\omega = \omega_1$ starting from $\boldsymbol \beta = \boldsymbol 0$, and set the first column of $\boldsymbol B$, $\boldsymbol b_1$, to the solution.
	\item For $l = 2, \dots, k_{\omega}$, solve for $\omega = \omega_{l}$ starting from the solution for $\omega_{l-1}$, $\boldsymbol b_{l-1}$, and set the $l$-th column of $\boldsymbol B$, $\boldsymbol b_{l}$, to the solution for $\omega_l$.
	\item Return the $p\times k_{\omega}$ solution array $\bs B$.
\end{enumerate}
\caption{Fixed $q$.}
\label{alg:fq}
\end{algorithm}
The second  computes a sequence of solutions for fixed $\omega$ for varying values of $q$.

\begin{algorithm}[H]
\begin{enumerate}
	\item Fix a sequence of $k_q$ decreasing $q$ values $\bs q = \left\{q_1, \dots, q_{k_q}\right\}$, where $q_1 = 2$. Let $\bs B$ refer to the $p\times l$ array of solutions.
	\item For $q = q_1$, set the first column of $\boldsymbol B$ to the closed-form solution $\boldsymbol b_1 = \left(\boldsymbol X'\boldsymbol X + \boldsymbol I\right)^{-1}\boldsymbol X'\boldsymbol y$.
	\item For $l = 2, \dots, k_q$, solve for $q = q_l$ starting from the solution for $q_{l-1}$, $\boldsymbol b_{l-1}$, and set the $l$-th column of $\boldsymbol B$, $\boldsymbol b_{l}$, to the solution for $q_l$.
	\item Return the $p\times k_q$ solution array $\bs B$.
\end{enumerate}
\caption{Fixed $\omega$.}
\label{alg:fo}
\end{algorithm}

We refer to Algorithms~\ref{alg:fq} and \ref{alg:fo} as warm start algorithms, as both repeatedly solve \eqref{eq:algre} starting from initial values obtained by solving a similar problem previously.

\section{Demonstrations}

To demonstrate the utility of Algorithms~\ref{alg:fq} and \ref{alg:fo}, we consider  \color{black} five simulation settings and \color{black} applications to five datasets of varying size and structure. For all \color{black} simulated and real \color{black} datasets, the response $\boldsymbol y$ and covariates $\boldsymbol X$ \color{black} are \color{black} centered and scaled. 

\color{black} In all simulation settings, we assume $\bs y = \bs A \bs V \bs \beta + \bs z$, where $\bs V$ and $\bs A$ are $n \times p$ matrix and $p\times p$ matrices, $\bs z$ is a $n\times 1$ noise vector,  and nonzero elements of $\bs \beta$, elements of $\bs V$, and elements of $\bs z$ are independent, identically distributed standard normal random variables. Specification of $n$, $p$, $\bs A$, and the number of nonzero elements of $\bs \beta$ depends on the simulation setting as described in Table~\ref{tab:simulations}. We simulate $4$ datasets per setting. \color{black}

\begin{table}[H]
\centering
\color{black}
\begin{tabular}{c|c|c|c|c} \hline
Simulation & $n$ & $p$ & $\frac{1}{p}\sum_{i = 1}^p \mathbbm{1}_{\left\{\beta_i = 0 \right\}}$ &  $\bs A$\\ \hline 
1& 100 & 1,000 & 1 & $\bs I_p$\\
2 (Sparse $\bs \beta$) & 100 & 1,000 & $0.1$ & $\bs I_p$ \\
3 (Correlated $\bs X$) & 100 & 1,000  & 1 & $0.75 \left(\bs 1_p \bs 1_p'\right)  + \left(1 - 0.75\right) \bs I_p$\\
4 & 500 & 1,000 & 1 & $\bs I_p$\\
5 & 100 & 2,000 & 1& $\bs I_p$\\ \hline
\end{tabular}
\caption{\color{black}Simulation settings used to demonstrate pathwise coordinate descent Algorithms~\ref{alg:fq} and \ref{alg:fo} for $\ell_q$ regression. A $p\times p$ identity matrix is denoted by $\bs I_p$.}
\label{tab:simulations}
\end{table}

\color{black}When implementing Algorithm~\eqref{eq:alg} for fixed $q$, we consider 20 values of $\omega$ that are equally spaced on the log-scale from $\omega^{(min)}_1$ to $10^{-20}$, where $\omega^{(min)}_1$  is determined from the data. When implementing Algorithm~\eqref{alg:fo} for fixed $\omega$, we consider $k_q = 20$ equally spaced values $\boldsymbol q = \left\{2, \dots, 0.1\right\}$. For each simulated dataset, we implement Algorithms~\ref{alg:fq} and \ref{alg:fo} and their cold start alternatives for $10$ randomly selected unique orderings of the $p$ covariates.\color{black}

\begin{figure}[h]
	\centering
	\includegraphics{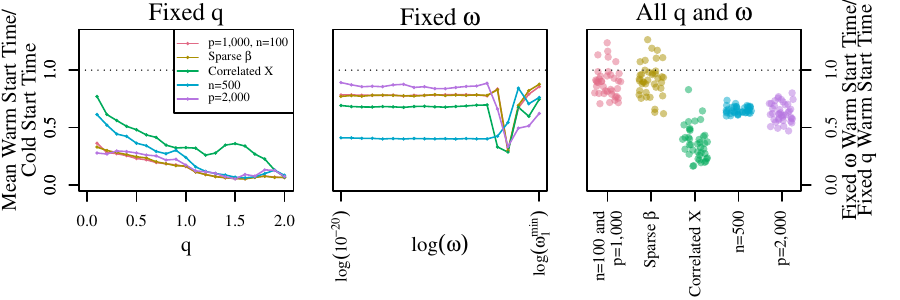}
	\caption{\color{black}The left and center panels show the mean ratio of time needed to complete Algorithms~\ref{alg:fq} and \ref{alg:fo} compared to their cold start alternatives across 100 randomly selected orderings of the $p$ covariates. The right panel shows the ratio of time needed to complete Algorithm~\ref{alg:fo} relative to Algorithm~\ref{alg:fq} for all $q$ and $\omega$ for each of the 100 randomly selected orderings of the $p$ covariates.}
	\label{fig:timingsim}
\end{figure}

\color{black}
Figure~\ref{fig:timingsim} depicts timing comparisons for Algorithms~\ref{alg:fq} and \ref{alg:fo} relative to cold start alternatives for simulated data. We compare Algorithm~\ref{alg:fq} to a cold start alternative that \color{black}minimizes \eqref{eq:algre} starting from $\boldsymbol \beta = \boldsymbol 0$, and we compare Algorithm~\ref{alg:fo} to a cold start alternative that \color{black}minimizes  \eqref{eq:algre} starting from $\boldsymbol \beta = (\boldsymbol X'\boldsymbol X + \boldsymbol I_p)^{-1}\boldsymbol X'\boldsymbol y$. 
In general, both warm start algorithms provide substantial timing gains relative to their cold start alternatives, especially when the number of covariates $p$ is large. The relative gains of Algorithm~\ref{alg:fq} compared to its cold start alternative are greater than the relative gains of Algorithm~\ref{alg:fo} compared to its cold start alternative, however Algorithm~\ref{alg:fo} is often faster than  Algorithm~\ref{alg:fq} when considering time needed to compute solutions for all $q$ and $\omega$, especially when covariates are correlated or the dimension is greater.
\color{black}

Table~\ref{tab:datasets} summarizes the number of observations $n$, the number of covariates $p$, and the average absolute correlation between covariates. Citations for where each dataset has previously appeared in the penalized regression literature are also provided. 

\begin{table}[H]
\centering
\begin{tabular}{c|c|c|c|c} \hline
Dataset & $n$ & $p$ & $\frac{1}{p\left(p - 1\right)}\sum_{i\neq j}\left|\text{cor}\left(\boldsymbol x_i, \boldsymbol x_j \right)\right|$ & Source \\ \hline 
Prostate & 97 & 8 & 0.295 & \cite{Tibshirani1996} \\
Diabetes & 442 & 64 & 0.150 &  \cite{Efron2004} \\
Glucose & 68 & 72  & 0.174 & \cite{Priami2015} \\
Housing & 506 & 104 & 0.360 & \cite{Polson2014} \\
Motif & 287 & 195 & 0.641 & \cite{Buhlmann2011}\\ \hline
\end{tabular}
\caption{Features of datasets used to demonstrate pathwise coordinate descent Algorithms~\ref{alg:fq} and \ref{alg:fo} for $\ell_q$ regression.}
\label{tab:datasets}
\end{table}

When implementing Algorithm~\eqref{eq:alg} for fixed $q$, we consider 20 \color{black} values of $\omega$ that are \color{black} equally spaced on the log-scale from $\omega^{(min)}_1$ to $10^{-7}$, where $\omega^{(min)}_1$ \color{black} is \color{black} determined from the data. \color{black} When implementing Algorithm~\eqref{alg:fo} for fixed $\omega$, we consider $k_q = 20$ equally spaced values $\boldsymbol q = \left\{2, \dots, 0.1\right\}$. For each dataset, we implement Algorithms~\ref{alg:fq} and \ref{alg:fo} and their cold start alternatives for \color{black}$100$ \color{black} randomly selected unique orderings of the $p$ covariates.

\begin{figure}[h]
	\centering
	\includegraphics{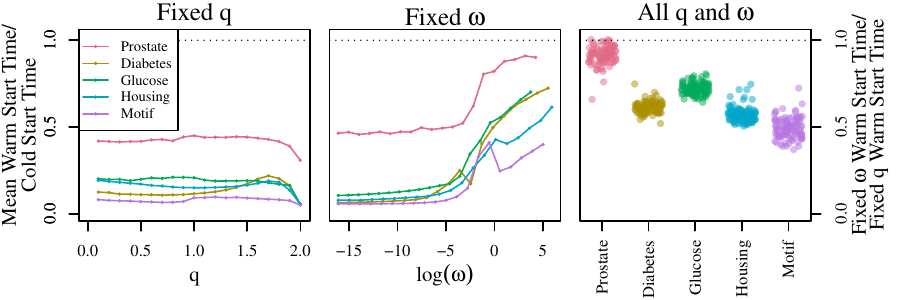}
	\caption{\color{black}The left and center panels show the mean ratio of time needed to complete Algorithms~\ref{alg:fq} and \ref{alg:fo} compared to their cold start alternatives across 100 randomly selected orderings of the $p$ covariates. The right panel shows the ratio of time needed to complete Algorithm~\ref{alg:fo} relative to Algorithm~\ref{alg:fq} for all $q$ and $\omega$ for each of the 100 randomly selected orderings of the $p$ covariates.}
	\label{fig:timing}
\end{figure}

Figure~\ref{fig:timing} depicts timing comparisons for Algorithms~\ref{alg:fq} and \ref{alg:fo} relative to cold start alternatives. We compare Algorithm~\ref{alg:fq} to a cold start alternative that \color{black}minimizes \color{black} \eqref{eq:algre} starting from $\boldsymbol \beta = \boldsymbol 0$, and we compare Algorithm~\ref{alg:fo} to a cold start alternative that \color{black}minimizes \color{black} \eqref{eq:algre} starting from $\boldsymbol \beta = (\boldsymbol X'\boldsymbol X + \boldsymbol I_p)^{-1}\boldsymbol X'\boldsymbol y$. 
\color{black}Again, both warm start algorithms provide substantial timing gains relative to their cold start alternatives, especially when the number of covariates $p$ is large. Perhaps due to the substantial correlations across covariates in all five of the real datasets, speed advantages of using Algorithm~\ref{alg:fo} over  Algorithm~\ref{alg:fq}  to compute solutions for all $q$ and $\omega$ are more pronounced. \color{black}
\color{black}To assess the extent to which speed advantages of Algorithm~\ref{alg:fo} are driven by performance when $q > 1$, which is less often of interest in practice, we have comparisons of a variation of Algorithm~\ref{alg:fo} that considers only $q \leq 1$ to its cold start alternative and Algorithm~\ref{alg:fq} in the appendix. We find that speed advantages of  Algorithm~\ref{alg:fo} are only slightly diminished.\color{black}

\begin{figure}[h]
	\centering
	\includegraphics{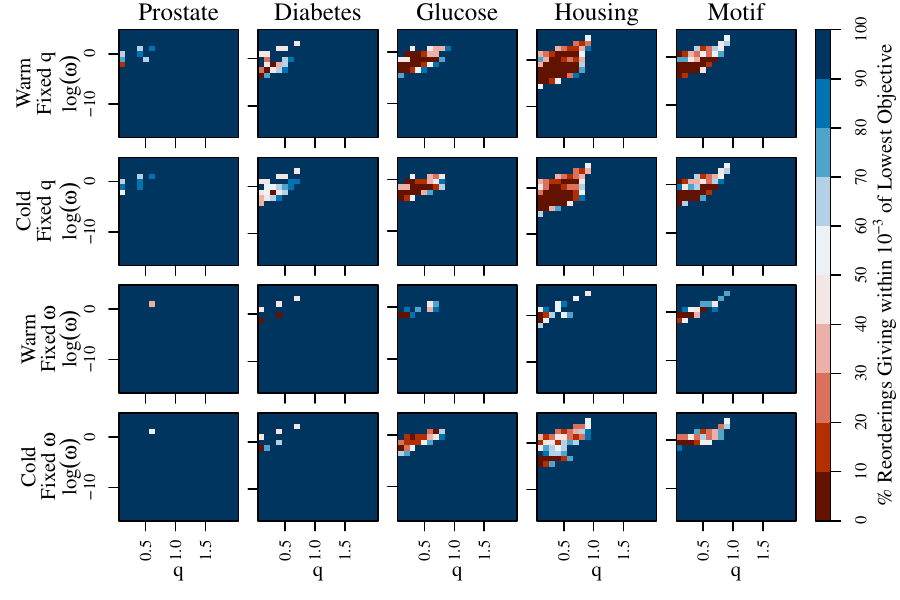}
	\caption{\color{black}For each dataset, algorithm, and pair of tuning parameter values $q$ and $\omega$, the proportion of random orderings of the covariates where the corresponding algorithm returns an objective function value within $10^{-3}$ of the lowest objective function value achieved by any of the remaining three algorithms using the same ordering. \color{black}Warm fixed $q$ corresponds to Algorithm~\ref{alg:fq}, cold fixed $q$ corresponds to Algorithm~\ref{alg:fq}'s cold start alternative, warm fixed $\omega$ corresponds to Algorithm~\ref{alg:fo}, and cold fixed $\omega$ corresponds to Algorithm~\ref{alg:fo}'s cold start alternative.}
	\label{fig:objective}
\end{figure}

A natural question given the favorable timing results for warm start algorithms shown in Figure~\ref{fig:timing} is whether or not timing gains come at the cost of poorer solutions \color{black} when $q < 1$ and \eqref{eq:alg} is nonconvex. \color{black} Figure~\ref{fig:objective} compares the rate at which each algorithm reaches an objective value within $10^{-3}$ of the lowest objective value obtained using the same ordering of covariates. \color{black} We do not observe that the timing gains associated with  Algorithms~\ref{alg:fq} and \ref{alg:fo} relative to their cold start algorithms come at the cost of poorer solutions. Algorithm~\ref{alg:fq} tends to provide comparable solutions to its cold start alternative for \color{black} all five datasets\color{black}. Algorithm~\ref{alg:fo} tends to provide comparable solutions to its cold start alternative for the prostate and diabetes datasets and better solutions for the glucose, housing, and motif datasets.  \color{black} In general, Algorithm~\ref{alg:fo} tends to provide the best solutions.\color{black} 

\section{Discussion}

In this paper, we have demonstrated that a new reparametrization of the $\ell_q$ penalized regression problem is well-suited to pathwise coordinate descent algorithms. 
There are several potential areas of improvement. 
\color{black} One is to explicitly examine which algorithms return solutions that satisfy the conditions provided for local optimality in \cite{Marjanovic2014} when $q < 1$, instead of considering which algorithms tend to return solutions with the lowest value of the objective function. \color{black}
\color{black}Another \color{black} is consideration of the number and spacing of values of $\omega$ and $q$ for Algorithms~\ref{alg:fq} and \ref{alg:fo}, respectively, which may determine the extent of Algorithms~\ref{alg:fq} and \ref{alg:fo}'s gains relative to cold start alternatives.
\color{black}A third \color{black} is consideration of the ordering of covariates in coordinate descent. It is known that flexibility with respect to the ordering of covariates is a specific advantage of coordinate descent methods, and it is possible that gains from Algorithms~\ref{alg:fq} and \ref{alg:fo} relative to their cold start alternatives could be enhanced by certain choices of covariate order \color{black} such as the ordering used by the active shooting algorithm described in \cite{Peng2009} and the orderings discussed in \color{black} \cite{Chartrand2016}. 
\color{black} A fourth is use of alternative methods for solving \eqref{eq:alg} for fixed values of $q$ and $\omega$, e.g.\ the local quadratic approximation approach of \cite{Fan2001}\color{black}, the modified Newton-Raphson approach for $1 < q < 2$ described in \cite{Fu1998}, \color{black} or other alternatives reviewed in \cite{Chartrand2016}. \color{black}
\color{black} Last, further work may consider the choice of a single solution to \eqref{eq:alg} when multiple exist. \color{black}

\addcontentsline{toc}{section}{References}
\singlespacing
\bibliography{library}

\begin{thebibliography}{}

\bibitem[\protect\citeauthoryear{Bühlmann and van~de Geer}{Bühlmann and
  van~de Geer}{2011}]{Buhlmann2011}
Bühlmann, P. and S.~van~de Geer (2011).
\newblock {\em Statistics for High-Dimensional Data: Methods, Theory and
  Applications}.
\newblock Springer.

\bibitem[\protect\citeauthoryear{Chartrand and Yin}{Chartrand and
  Yin}{2016}]{Chartrand2016}
Chartrand, R. and W.~Yin (2016).
\newblock Nonconvex sparse regularization and splitting algorithms.

\bibitem[\protect\citeauthoryear{Efron, Hastie, Johnstone, and
  Tibshirani}{Efron et~al.}{2004}]{Efron2004}
Efron, B., T.~Hastie, I.~Johnstone, and R.~Tibshirani (2004).
\newblock Least angle regression.
\newblock {\em The Annals of Statistics\/}~{\em 32}, 407--499.

\bibitem[\protect\citeauthoryear{Fan and Li}{Fan and Li}{2001}]{Fan2001}
Fan, J. and R.~Li (2001).
\newblock Variable selection via nonconcave penalized likelihood and its oracle
  properties.
\newblock {\em Journal of the American Statistical Association\/}~{\em 96},
  1348--1360.

\bibitem[\protect\citeauthoryear{Frank and Friedman}{Frank and
  Friedman}{1993}]{Frank1993}
Frank, I.~E. and J.~H. Friedman (1993).
\newblock A statistical view of some chemometrics regression tools.
\newblock {\em Technometrics\/}~{\em 35}, 109.

\bibitem[\protect\citeauthoryear{Friedman, Hastie, Höfling, and
  Tibshirani}{Friedman et~al.}{2007}]{Friedman2007}
Friedman, J.~H., T.~Hastie, H.~Höfling, and R.~Tibshirani (2007).
\newblock Pathwise coordinate optimization.
\newblock {\em Annals of Applied Statistics\/}~{\em 1}, 302--332.

\bibitem[\protect\citeauthoryear{Fu}{Fu}{1998}]{Fu1998}
Fu, W.~J. (1998).
\newblock Penalized regressions: The bridge versus the lasso.
\newblock {\em Journal of Computational and Graphical Statistics\/}~{\em 7},
  397--416.

\bibitem[\protect\citeauthoryear{Griffin and Hoff}{Griffin and
  Hoff}{2020}]{Griffin2017c}
Griffin, M. and P.~D. Hoff (2020).
\newblock Testing sparsity inducing penalties.
\newblock {\em Journal of Computational and Graphical Statistics\/}~{\em 29},
  1--12.

\bibitem[\protect\citeauthoryear{Huang, Horowitz, and Ma}{Huang
  et~al.}{2008}]{Huang2008}
Huang, J., J.~L. Horowitz, and S.~Ma (2008).
\newblock Asymptotic properties of bridge estimators in sparse high-dimensional
  regression models.
\newblock {\em The Annals of Statistics\/}~{\em 36}, 587--613.

\bibitem[\protect\citeauthoryear{Marjanovic and Solo}{Marjanovic and
  Solo}{2014}]{Marjanovic2014}
Marjanovic, G. and V.~Solo (2014).
\newblock lq sparsity penalized linear regression with cyclic descent.
\newblock {\em IEEE Transactions on Signal Processing\/}~{\em 62}, 1464--1475.

\bibitem[\protect\citeauthoryear{Mazumder, Friedman, and Hastie}{Mazumder
  et~al.}{2011}]{Mazumder2011}
Mazumder, R., J.~H. Friedman, and T.~Hastie (2011).
\newblock Sparsenet: Coordinate descent with nonconvex penalties.
\newblock {\em Journal of the American Statistical Association\/}~{\em 106},
  1125--1138.

\bibitem[\protect\citeauthoryear{Peng, Wang, Zhou, and Zhu}{Peng
  et~al.}{2009}]{Peng2009}
Peng, J., P.~Wang, N.~Zhou, and J.~Zhu (2009, 6).
\newblock Partial correlation estimation by joint sparse regression models.
\newblock {\em Journal of the American Statistical Association\/}~{\em 104},
  735--746.

\bibitem[\protect\citeauthoryear{Polson, Scott, and Windle}{Polson
  et~al.}{2014}]{Polson2014}
Polson, N.~G., J.~G. Scott, and J.~Windle (2014).
\newblock The bayesian bridge.
\newblock {\em Journal of the Royal Statistical Society. Series B: Statistical
  Methodology\/}~{\em 76}, 713--733.

\bibitem[\protect\citeauthoryear{Priami and Morine}{Priami and
  Morine}{2015}]{Priami2015}
Priami, C. and M.~J. Morine (2015).
\newblock {\em Analysis of Biological Systems}.
\newblock Imperial College Press.

\bibitem[\protect\citeauthoryear{Tibshirani}{Tibshirani}{1996}]{Tibshirani1996}
Tibshirani, R. (1996).
\newblock Regression shrinkage and selection via the lasso.
\newblock {\em Journal of the Royal Statistical Society: Series B (Statistical
  Methodology)\/}~{\em 58}, 267--288.

\end{thebibliography}
\bibliographystyle{chicago}
\pagebreak

\begin{appendix}
\noindent \emph{Proof of Theorem 2.1: } 
 Taking the first derivative of $\alpha\left(\omega, q\right)$ with respect to $\omega$, we obtain
\begin{align*}
\frac{\partial \alpha\left(\omega, q\right)}{\partial \omega} &= \left(2\left(1 - q\right)\right)^{\frac{q - 1}{2-q}}\left(2 - q\right) q^{\frac{1}{q-2}} > 0,
\end{align*} thus sparsity of the mode-thresholding function is increasing in $\omega$,  i.e. if $h\left(\omega, q; b\right) = 0$ and $\omega' > \omega$ implies $h\left(\omega', q; b\right) = 0$.  \\

\noindent \emph{Proof of Theorem 2.2: }
When $\omega = \left|b\right|/2$, the mode thresholding function is nonzero for $q$ satisfying $\left|b\right| > \alpha \left(\omega = \left|b\right|/2, q\right)$, which corresponds to $q$ satisfying Equation~\eqref{eq:qcond}. Nonzero mode thresholding function values $h\left(\omega = \left|b\right|/2, q; b\right) = \text{sign}\left(b\right)\phi\left(\omega = \left|b\right|/2, q\right)$ have absolute value $\phi\left(\omega = \left|b\right|/2, q\right)$, which is the largest of at most two values satisfying Equation~\eqref{eq:phire} for $\omega = \left|b\right|/2$.
Regardless of $q$, $\phi\left(\omega = \left|b\right|/2, q\right) = \left|b\right|/2$ satisfies Equation~\eqref{eq:phire} for $\omega = \left|b\right|/2$. We prove that $\phi\left(\omega = \left|b\right|/2, q\right)$ is the largest value that satisfies this equation by contradiction. Suppose that there is a larger value $\phi\left(\omega = \left|b\right|/2, q\right) = \left(1 + \epsilon\right)  \left|b\right|/2$ for $\epsilon > 0$ that satisfies Equation~\eqref{eq:phire} for $\omega = \left|b\right|/2$. We have: 
\begin{align*}
\phi\left(\omega = \left|b\right|/2, q\right) + \left(\frac{\left|b\right|}{2} \right)\left(\frac{\phi\left(\omega = \left|b\right|/2, q\right)}{\left|b\right|/2}\right)^{q-1} &=  \left(1 + \epsilon + \left(1 + \epsilon\right)^{q - 1}\right) \left(\frac{\left|b\right|}{2} \right) \\
&>  \left(1 + \epsilon + \left(1 + \epsilon\right)^{-1}\right) \left(\frac{\left|b\right|}{2} \right) \\
&= \left(2 + \frac{\epsilon^2}{1 + \epsilon}\right) \left(\frac{\left|b\right|}{2} \right) > \left|b\right|.
\end{align*}
Thus, $\phi\left(\omega = \left|b\right|/2, q\right) = \left(1 + \epsilon\right)  \left|b\right|/2$ fails to satisfy Equation~\eqref{eq:phire} for $\omega = \left|b\right|/2$ when $\epsilon > 0$, $\phi\left(\omega = \left|b\right|/2, q\right) = \left|b\right|/2$ is the largest value that satisfies Equation~\eqref{eq:phire}, and $h\left(\omega = \left|b\right|/2, q; b\right) = \text{sign}\left(b\right)\left|b\right|/2 = b/2$. \\

\noindent \emph{Proof of Theorem 2.3: } Suppose that $\omega' < \omega$. For all $\beta$ including $\beta =h\left(\omega', q; b\right)$, $h\left(\omega, q; b\right)$ satisfies
\begin{align}\label{eq:hmax}
\frac{1}{2}\left(h\left(\omega, q; b\right) - b\right)^2 + \left(\frac{\omega^{2-q}}{q}\right)\left|h\left(\omega, q; b\right) \right|^q \leq \frac{1}{2}\left(\beta- b\right)^2 + \left(\frac{\omega^{2-q}}{q}\right)\left|\beta \right|^q.
\end{align}
Likewise, for all $\beta$ including $\beta =h\left(\omega, q; b\right)$, $h\left(\omega', q; b\right)$ satisfies
\begin{align}\label{eq:hmaxp}
\frac{1}{2}\left(h\left(\omega', q; b\right) - b\right)^2 + \left(\frac{\omega'^{2-q}}{q}\right)\left|h\left(\omega', q; b\right) \right|^q \leq \frac{1}{2}\left(\beta- b\right)^2 + \left(\frac{\omega'^{2-q}}{q}\right)\left|\beta \right|^q.
\end{align} 
Using the facts that the inequality in Equation~\eqref{eq:hmax} holds for $\beta =h\left(\omega', q; b\right)$ and the inequality in Equation~\eqref{eq:hmaxp} holds for $\beta =h\left(\omega, q; b\right)$, adding both inequalities together, and simplifying yields
\begin{align*}
\left(\frac{\omega^{2-q} - \omega'^{2-q}}{q}\right)\left|h\left(\omega, q; b\right) \right|^q \leq \left(\frac{\omega^{2-q} - \omega'^{2-q}}{q}\right)\left|h\left(\omega', q; b\right) \right|^q \implies \left|h\left(\omega, q; b\right) \right| \leq \left|h\left(\omega', q; b\right) \right|.
\end{align*} \\

\noindent \emph{Proof of Theorem 2.4: } 
Taking the first derivative of $\alpha\left(\omega, q\right)$ with respect to $q$, we obtain 
\begin{align*}
\frac{\partial \alpha\left(\omega, q\right)}{\partial q} = \alpha\left(\omega, q\right)\left(\frac{q\text{log}\left(2\right)+ q\text{log}\left(1 - q\right)+q - q \text{log}\left(q\right) - 2}{\left(q - 2\right)^2q}\right).
\end{align*}
Because $q > 0$, and $\alpha\left(\omega, q\right) > 0$ for $0 < q \leq 1$, the sign of this derivative depends strictly on the numerator of the second term, which can be bounded above by numerically maximizing $q\text{log}\left(2\right)+ q\text{log}\left(1 - q\right)+q - q \text{log}\left(q\right) - 2$ over $0 < q \leq 1$,
\begin{align*} 
q\text{log}\left(2\right)+ q\text{log}\left(1 - q\right)+q - q \text{log}\left(q\right) - 2 \leq -1.147.
\end{align*} 
Thus, $\frac{\partial \alpha\left(\omega, q\right)}{\partial q} < 0$ and sparsity of the mode-thresholding function is nested in $q$, i.e. if $h\left(\omega, q; b\right) = 0$ and $q' < q$ implies $h\left(\omega, q'; b\right) = 0$. \\

\noindent \emph{Proof of Theorem 2.5: }  For fixed $\omega > 0$, $\left|h\left(\omega, q; b\right)\right| > 0$ for $q \geq 1$ and $q < 1$ that satisfy:
\begin{align*}
\left(\frac{1}{2-q}\right) \left(2\left(1 - q\right)\right)^{\frac{1 - q}{2-q}}q^{\frac{1}{2-q}} > \omega/\left|b\right|.
\end{align*}
Recognizing that $\left(\frac{1}{2-q}\right) \left(2\left(1 - q\right)\right)^{\frac{1 - q}{2-q}}q^{\frac{1}{2-q}} = \alpha\left(\omega = 1, q\right)^{-1}$ and recalling the fact that $\alpha\left(\omega, q\right)$ is strictly decreasing in $q$ for fixed $\omega$, as shown in the proof of Theorem 2.2, the constraint  $\left(\frac{1}{2-q}\right) \left(2\left(1 - q\right)\right)^{\frac{1 - q}{2-q}}q^{\frac{1}{2-q}} > \omega/\left|b\right|$ defines an interval $\left[\tilde{q}_{\omega/\left|b\right|}, 1\right)$ for values of $q < 1$ that yield nonzero solutions $\left|h\left(\omega, q; b\right)\right| > 0$.

It follows from the fact that the function , that the function $\left(\frac{1}{2-q}\right) \left(2\left(1 - q\right)\right)^{\frac{1 - q}{2-q}}q^{\frac{1}{2-q}} = \alpha\left(\omega = 1, q\right)^{-1}$ is strictly increasing in $q$. 
Thus, if $q' > q$ and $\omega  < \left(\frac{\left|b\right|}{2-q}\right) \left(2\left(1 - q\right)\right)^{\frac{1 - q}{2-q}}q^{\frac{1}{2-q}}$ then $\omega   < \left(\frac{\left|b\right|}{2-q'}\right) \left(2\left(1 - q'\right)\right)^{\frac{1 - q'}{2-q'}}q'^{\frac{1}{2-q'}}$. Accordingly, if $q' > q$ then $\left|h\left(\omega, q; b\right)\right| > 0$ and $\left|h\left(\omega, q'; b\right)\right| > 0$.

For all $\beta$ including $\beta =h\left(\omega, q'; b\right)$, $h\left(\omega, q; b\right)$ satisfies
\begin{align}\label{eq:hmax2}
\frac{1}{2}\left(h\left(\omega, q; b\right) - b\right)^2 + \left(\frac{\omega^{2-q}}{q}\right)\left|h\left(\omega, q; b\right) \right|^q \leq \frac{1}{2}\left(\beta- b\right)^2 + \left(\frac{\omega^{2-q}}{q}\right)\left|\beta \right|^q.
\end{align}
Likewise, for all $\beta$ including $\beta =h\left(\omega, q; b\right)$, $h\left(\omega, q'; b\right)$ satisfies
\begin{align}\label{eq:hmaxp2}
\frac{1}{2}\left(h\left(\omega, q'; b\right) - b\right)^2 + \left(\frac{\omega^{2-q'}}{q'}\right)\left|h\left(\omega, q'; b\right) \right|^{q'} \leq \frac{1}{2}\left(\beta- b\right)^2 + \left(\frac{\omega^{2-q'}}{q'}\right)\left|\beta \right|^{q'}.
\end{align}

Using the facts that the inequality in Equation~\eqref{eq:hmax2} holds for $\beta =h\left(\omega, q'; b\right)$ and the inequality in Equation~\eqref{eq:hmaxp2} holds for $\beta =h\left(\omega, q; b\right)$, adding both inequalities together, and simplifying yields
\footnotesize
\begin{align}\label{eq:nestq}
  \left(\frac{1}{q}\right)\left|h\left(\omega, q; b\right)/\omega \right|^q  - \left(\frac{1}{q'}\right)\left|h\left(\omega, q; b\right)/\omega \right|^{q'} \leq \left(\frac{1}{q}\right)\left|h\left(\omega, q'; b\right)/\omega \right|^q -  \left(\frac{1}{q'}\right)\left|h\left(\omega, q'; b\right)/\omega \right|^{q'} 
  \end{align}
  \normalsize
  This becomes a question about the behavior of the function $x^q/q - x^{q'}/q'$ with respect to $x$; is it increasing or decreasing in $x$ for $q' > q$? If it is increasing in $x$, then the inequality above implies that $\left|h\left(\omega, q'; b\right)\right| \geq \left|h\left(\omega, q; b\right)\right|$. If it is decreasing in $x$, then the inequality above implies that $ \left|h\left(\omega, q; b\right)\right| \geq \left|h\left(\omega, q'; b\right)\right|$.
  
 The derivative of the function $x^q/q - x^{q'}/q'$ with respect to $x$ is 
 \begin{align*}
 x^{q-1} - x^{q' - 1} = x^{q-1} \left(1 - x^{q' - q} \right).
 \end{align*} 
 This is positive when $x < 1$, equal to $0$ when $x = 1$, and negative when $x > 1$. Recalling \eqref{eq:nestq}, we are interested in the sign of the derivative of the function $x^q/q - x^{q'}/q'$ for 
 \begin{align*}
 \text{min}\left\{\frac{\left|h\left(\omega, q; b\right)\right|}{\omega}, \frac{\left|h\left(\omega, q'; b\right)\right|}{\omega}\right\} \leq x \leq \text{max}\left\{\frac{\left|h\left(\omega, q; b\right)\right|}{\omega}, \frac{\left|h\left(\omega, q'; b\right)\right|}{\omega}\right\}.
 \end{align*}

Applying Theorems 2.3 and 2.4, we have that for any $0 < q \leq 2$,  if $\left|b\right|/2 \geq \omega$, then $\left|h\left(\omega, q, ; b\right)\right| \geq \left|b\right|/2$. It follows that $\left|h\left(\omega, q, ; b\right)\right|/\omega \geq 1$ and $\left|h\left(\omega, q', ; b\right)/\omega\right| \geq 1$. Thus, when $\left|b\right|/2 \geq \omega$ the sign of the derivative of the function $x^q/q - x^{q'}/q'$ is of interest for
 \begin{align*}
1 \leq x \leq  \text{max}\left\{\frac{\left|h\left(\omega, q; b\right)\right|}{\omega}, \frac{\left|h\left(\omega, q'; b\right)\right|}{\omega}\right\} ,
 \end{align*}
 and thus always negative. It follows that $ \left|h\left(\omega, q; b\right)\right| \geq \left|h\left(\omega, q'; b\right)\right|$ when $\left|b\right|/2 \geq \omega$.

Applying Theorems 2.3 and 2.4 again, we have that for any $0 < q \leq 2$, if $\left|b\right|/2 \leq \omega$, then $\left|b\right|/2 \geq \left|h\left(\omega, q;  b\right)\right|$. It follows that $\left|h\left(\omega, q;  b\right)/\omega\right| \leq 1$ and $\left|h\left(\omega, q';  b\right)/\omega\right| \leq 1$. Thus, when $\left|b\right|/2 \leq \omega$ the sign of the derivative of the function $x^q/q - x^{q'}/q'$ is of interest for
 \begin{align*}
 \text{min}\left\{\frac{\left|h\left(\omega, q; b\right)\right|}{\omega}, \frac{\left|h\left(\omega, q'; b\right)\right|}{\omega}\right\} \leq x \leq 1,
 \end{align*}
 and thus always positive. It follows that $\left|h\left(\omega, q'; b\right)\right| \geq \left|h\left(\omega, q; b\right)\right|$ when $\left|b\right|/2 \leq \omega$. \\

\clearpage
\noindent \emph{\color{black}Timing Comparisons for $q \leq 1$:}

\begin{figure}[h]
	\centering
	\includegraphics{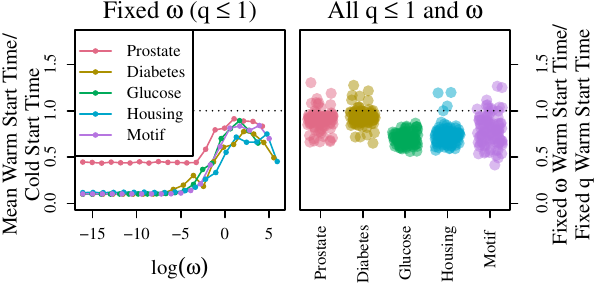}
	\caption{\color{black}The left panel shows the mean ratio of time needed to complete a variation of Algorithm \ref{alg:fo} that starts from $q = 1$ using the solution for $q = 2$ as a starting value compared to its cold start alternatives across 100 randomly selected orderings of the $p$ covariates for fixed $q \leq 1$. The right panel shows the ratio of time needed to complete this variation of Algorithm~\ref{alg:fo} relative to Algorithm~\ref{alg:fq} for all $q \leq 1$ and $\omega$ for each of the 100 randomly selected orderings of the $p$ covariates.}
	\label{fig:timingsim}
\end{figure}

\clearpage
\noindent \emph{\color{black}Objective Comparisons for Simulations:}

\begin{figure}[h]
	\centering
	\includegraphics{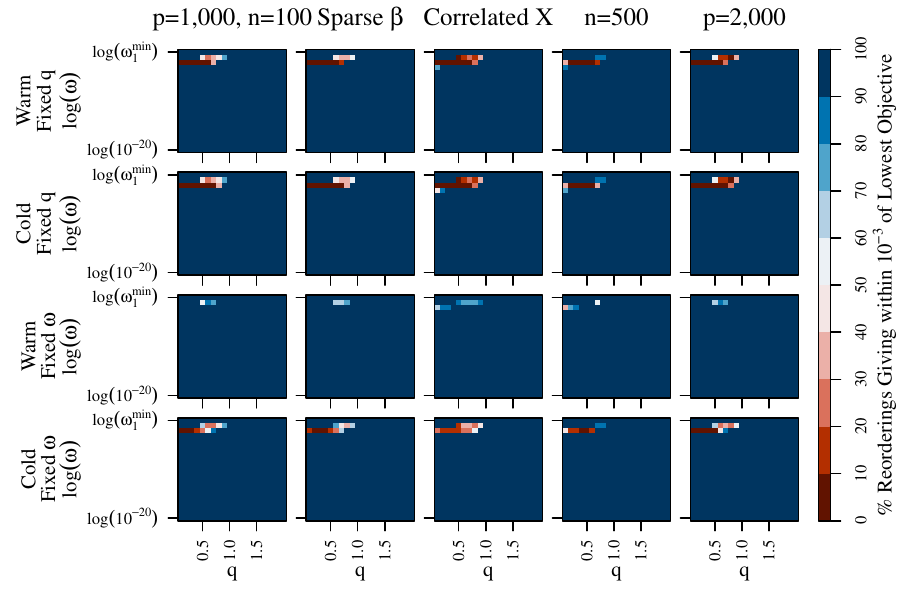}
	\caption{\color{black}For each simulation setting, algorithm, and pair of tuning parameter values $q$ and $\omega$, the proportion of random orderings of the covariates where the corresponding algorithm returns an objective function value within $10^{-3}$ of the lowest objective function value achieved by any of the remaining three algorithms using the same ordering. Warm fixed $q$ corresponds to Algorithm~\ref{alg:fq}, cold fixed $q$ corresponds to Algorithm~\ref{alg:fq}'s cold start alternative, warm fixed $\omega$ corresponds to Algorithm~\ref{alg:fo}, and cold fixed $\omega$ corresponds to Algorithm~\ref{alg:fo}'s cold start alternative.}
	\label{fig:objectivesim}
\end{figure}

\end{appendix}
	
\end{document}